\documentclass[preprint,12pt]{elsarticle}
\usepackage{amssymb}
\usepackage{mathtools}
\usepackage{enumerate}
\usepackage[margin=3.0cm]{geometry}
\usepackage{comment}
\usepackage{color}
\usepackage{fancybox}
\usepackage{multirow}
\usepackage{tabu}

\usepackage{subfigure}

\begin{document}
\begin{frontmatter}
\title{Engineering solitons and breathers in a deformed ferromagnet: Effect of localised inhomogeneities}
\author[label1]{M. Saravanan}
\ead{saravanan\_manickam@yahoo.com}
\address[label1]{Department of Physics, Faculty of Engineering and Technology, SRM Institute of Science and Technology, Ramapuram Campus, Ramapuram, Chennai- 600 089, Tamilnadu, India.}
\author[label2]{A. Arnaudon}
\address[label2]{Department of Mathematics, Imperial College, London SW7 2AZ, UK.}
\begin{abstract}
We investigate the soliton dynamics of the electromagnetic wave propagating in an inhomogeneous or deformed ferromagnet. The dynamics of magnetization and the propagation of electromagnetic waves are governed by the Landau-Lifshitz-Maxwell (LLM) equation, a certain coupling between the Landau-Lifshitz and Maxwell's equations. In the framework of multiscale analysis, we obtain the perturbed integral modified KdV (PIMKdV) equation. 
Since the dynamic is governed by the nonlinear integro-differential equation, we rely on numerical simulations to study the interaction of its mKdV solitons with various types of inhomogeneities. 
Apart from simple one soliton experiments with periodic or localised inhomogeneities, the numerical simulations revealed an interesting dynamical scenario where the collision of two solitons on a localised inhomogeneity create a bound state which then produces either two separated solitons or a mKdV breather. 
\end{abstract}
\begin{keyword}
Reductive perturbation method, inhomogeneous ferromagnet, electromagnetic wave propagation, mKdV breathers 
\end{keyword}
\end{frontmatter}
\section{Introduction}

The problem of nonlinear excitations in ferromagnetic models has been under extensive investigations for many years due to its wide range of applications in real material systems \cite{1,2,3,4}. For instance, ferromagnetic materials with different interactions have had a lot of importance in the context of data storage and allowed a faster coding of binary information. With these types of exchange interactions in the material, a localization phenomenon can be used to improve on these technological issues. Based on the availability of various nonlinear mechanisms in different physical fields, Ref \cite{5} has shown the importance of soliton dynamics for understanding physical phenomena and designing new experiments. Remarkable experiments in the past decade \cite{6,7} used the propagation of electromagnetic (EM) waves in ferromagnetic mediums to demonstrate that magnetic field component of this EM wave could serve as a better option for the storage technology, using the fact that the magnetization can be manipulated to be aligned with a certain direction. This process is called magnetization reversal or switching and is possible only because the magnetization is localised as a soliton structure in both the space and time domain. 

These qualitative features of the effect of EM waves on the magnetization dynamics can be understood theoretically by solving the associated nonlinear equations \cite{8,9}. 
The soliton propagation in a ferromagnetic medium under the effect of EM wave was first studied by Nakata \cite{8} using a multiscale approach on the celebrated Landau-Lifshitz equation coupled with the Maxwell equation. The magnetization dynamics in terms of solitary waves with long wavelengths is governed by the modified KdV equation. Extended studies were carried out by Leblond \cite{10,11,12,13,14,15} who developed similar theories for several such applications. 
In these models, the dispersion relation has two possible phase velocities for the solitons,  one case is studied in \cite{8} and the other propagation mode was studied by Leblond in \cite{11} and corresponds to KdV equation.  In case of ferromagnetic slabs, a short solitary wave has been derived in \cite{13} and the excitations are classified in polariton range. The background instability is suppressed by narrowing the slab and corresponds to the magnetization reversal. 

When damping is included in the system, the nonlinear modulation and the solitons are governed by nonlinearity, dispersion and damping \cite{14}. For small values of damping, the magnetization is governed by nonlinear Schr${\ddot{\textmd{o}}}$dinger (NLS) equation and the solitons cancel the effect due to damping. For large values of the damping, the wave decreases exponentially and no nonlinear modulation can occur and the dynamic is described by a perturbed NLS equation. In the three-dimensional case, the expected localised electromagnetic pulse was obtained in the focusing and defocusing cases from Davey-Stewartson (DS) equation admitting certain integrability conditions, as reported in \cite{15}. The reduction to (2+1)-dimensional system is not integrable through the IST method for the hyperbolic-hyperbolic and elliptic-elliptic defocusing case. Corrections to the above model in the one-dimensional case have been obtained by taking into account Heisenberg exchanges, uniaxial anisotropies, the antiferromagnetic character of the medium and effects of damping due to the presence of free charges \cite{16,17,18}. In these models, the propagation dynamics is described by the usual NLS family of equations. The effect of perturbation due to the presence of free charges in a conducting ferromagnetic medium \cite{17} shows structural perturbations on the soliton, such as for example in \cite{14}. 

In the studies we just mentioned, the focus was mainly on the magnetization of the medium and not on the wave evolution. However, waves should not be excluded when an exchange coupling is introduced in the ferromagnetic medium. In this setting, nonlinear excitations in classical models with significant exchange couplings have been studied widely in the past for the existence of solitons \cite{19,20,21,22,23,24,25,26,27,28,29,30}. These investigations are devoted to the excitations and soliton solutions that are affected by the EM waves. Hence, it is necessary to exploit the excitations including the higher order exchange interactions. 

Recently, the present author investigated the helimagnetic system in analogy to cholesteric liquid crystal and showed that EM waves are modulated in the form of solitons described by generalized derivative NLS equation \cite{31}. 
In this letter, the inhomogeneous or site-dependent exchange model is considered for the study of soliton dynamics when coupled with the EM wave propagation. 

\subsection*{Structure of the paper}
In section \ref{sec-2} of this paper, we present the model and discuss the inhomogeneity of magnetic materials. 
In section \ref{sec-3} the dynamical system is reduced to the perturbed integral modified KdV equation using the multiscale analysis. Section \ref{sec-4} is devoted to solving the PIMKdV equation using numerical simulations and several effects of inhomogeneities are discussed, in particular periodic, and localised inhomogeneities, with the new process of emergence of breathers from soliton collisions on a wall.  
The summary of the investigation is presented in section \ref{sec-5}. 

\section{Inhomogeneous magnetic system}\label{sec-2}

\subsection{The Maxwell-Landau equation}

We use a coupling between the classical Landau-Lifshitz model for the magnetization density function ${\bf{M}}(x,t)=(M^{x},M^{y},M^{z})$ and Maxwell's equations for the propagation of EM waves in ferromagnetic  chains, see \cite{32}. This results in the celebrated Maxwell-Landau (ML) model taking into account the Heisenberg exchange couplings and inhomogeneities while neglecting anisotropies, given by
\begin{align}
\frac{\partial{\bf{M}}}{\partial t}&={\bf{M}}\wedge \big[J(f{\bf{M}}_{xx}+f_{x}{\bf{M}}_{x})+\gamma {\bf{H}}\big],\label{1}\\
 \frac{1}{c^{2}}\frac{\partial^{2}}{\partial t^{2}}({\bf{H}}+{\bf{M}}) &= -\nabla(\nabla\cdot{\bf{H}})+\nabla^{2}{\bf{H}}\, ,\label{2}
\end{align}
where the constant $J$ is the exchange integral, $\gamma$ is the gyromagnetic ratio and the function $f$ captures the inhomogeneities. In general $f$ is a function of both space and time, such as $f=\mu_{1}(t)x+\nu_{1}(t)$ but the physical interpretation as a model for a ferromagnet is less clear  for time-dependent functions. $f$ is therefore taken as a time-independent function \cite{33}. 
Finally, $c^{2}=1/\sqrt{\mu_{0}\varepsilon_{0}}$, where $\mu_{0}$ is the magnetic permeability and $\varepsilon_{0}$ is the dielectric constant of the medium. The external magnetic field is coupled via the magnetic field component ${\bf{H}}$ of the EM wave. 
Since this model is inhomogeneous we only focus here on the one-dimensional study of the above equations \eqref{1} and \eqref{2} and leave the study of this higher dimensional equation for future works.

\subsection{On inhomogeneities}

Inhomogeneities otherwise known as deformations in a system may either be due to external fields or to the presence of defects, voids and gaps in the material. 
In the first case, inhomogeneities arise when a ferromagnetic medium lying in the $(x-y)$ plane is magnetized either in the longitudinal axis say $x$ direction or transverse axis say $y$ direction.
When saturated along the longitudinal axis, a medium exhibits a homogeneous effective field ${\bf{H}}_{\textmd{int}}$ but when magnetized in the transverse axis, the induced effective field ${\bf{H}}_{\textmd{int}}$ is inhomogeneous, see \cite{34}. These types of deformations are called field deformations. 
In the other case, a site-dependent function $f$ is introduced into the Hamiltonian to model the lattice defects such that the corresponding exchange integral is bond dependent, see for example \cite{35}. These lattice defects introduce a lattice distortion thereby leading the material to a deformed one. 
Such a dependence can occur if (a) the distance between neighbouring atoms varies along the chain, hence altering the overlap of electronic wavefunctions (assumed to be identical at all sites), or (b), if the wavefunction itself varies from site to site, even for equally spaced atoms. 
As examples of the case (a), we can mention charge transfer complexes TCNQ \cite{36} where the inhomogeneity function $f$ acting as a coefficient of exchange interactions is a set of random variables. 
Another example arises in organo-metallic insulators TTF-bisdithiolenes \cite{37} for which $f$ randomly alternates between two values along the chain. 
A chain system is natural for modelling inhomogeneities due to defects, although it may still be applicable in the case of weakly disordered systems with peaked wavefunctions such that a small change in the lattice constant causes a relatively large change in the atom overlap. 
These inhomogeneities can be modelled in the effective Hamiltonian for a one-dimensional magnetic insulator placed in a weak, static, inhomogeneous electric field or by the introduction of imperfections (impurities or organic complexes) in the vicinity of a bond to alter the electronic wavefunctions without causing appreciable lattice distortions. 
By gradually changing the concentration of impurities along the chain, it is possible to engineer a controlled inhomogeneity function $f$. 

\section{Dynamics of Maxwell-Landau model}\label{sec-3}

\subsection{Scaling}

We will now use the approach of multiscale analysis to reduce the coupled vector equations \eqref{1} and \eqref{2} to a nonlinear equation for a scalar field $u(x,t)$. 
The multiscale analysis or the reductive perturbation method is a generalized asymptotic analysis method for solving the Maxwell-Landau model by reducing it to soliton equations and possibly integrable equations \cite{16,17,18,38}. 
We first introduce the following slow space and time variables, depending on a small parameter $\varepsilon$ and two exponents $m$ and $n$, 
\begin{align}
\zeta=\varepsilon^{m}(x-Vt)\,, \qquad \tau=\varepsilon^{n}t\, .\label{4}
\end{align}
The slow space variable $\zeta$ describes the shape of the pulse propagating at speed $V$ and the time variable $\tau$ is the time evolution of this pulse during the propagation. The parameter $\varepsilon$ measures the weakness of the perturbation effect. 

This scaling was first introduced in the one dimension by  Gardner and Morikawa \cite{39} and later developed by Leblond \cite{15} for the (3+1) dimensional reduction.  
This scaling is also effective in studying anisotropic magnetic materials for soliton excitations, see \cite{40,41}. 

\subsection{Perturbation scheme} 

In connection to the theory of ferromagnetic resonance experiments \cite{42}, we assume that we model a sample immersed in a high intensity external magnetic field ${\bf{H}}_{\textmd{ext}}$ such that the sample is magnetized to its saturation limit.
This assumption avoids the creation of domain walls which would require more care in the derivation of the reduced equation and the sample is free from the effects due to its geometry such as shape anisotropies. 
Under this assumption, we seek a reduction of the coupled system \eqref{1} and \eqref{2} through the multiscale perturbation from the expansion  
\begin{align}
{\bf{M}}=\sum_{i=0}^{\infty}\varepsilon^{i}{\bf{M}}_{i} \quad \mathrm{and}\quad 
{\bf{H}}=\sum_{i=0}^{\infty}\varepsilon^{i}{\bf{H}}_{i}\, . \label{3}
\end{align}

We call the zeroth order term ${\bf{H}}_{0}$ the external field despite the fact that it differs from the applied external field ${\bf{H}}_{\textmd{ext}}$ as it is created by ${\bf{H}}_{\textmd{ext}}$. ${\bf{M}}_{0}$ will then be the magnetization density lined up with this field ${\bf{H}}_{0}$. 
The expansion of equation \eqref{3} is thus considered for perturbations around these fields (${\bf{M}}_{0}$,~${\bf{H}}_{0}$). 

\subsection{Nonlinear equation}

Gardner and Morikawa \cite{39} combined the scaling \eqref{4} with the perturbation expansion \eqref{3} for the components of  ${\bf{M}}$ and ${\bf{H}}$ in the aim of describing the nonlinear asymptotic behaviour of the original system.
The fractional choices of $m$ and $n$ are in general used for reducing to a scalar nonlinear equation of Schrodinger type but here, we instead used $m=1$ and $n=3$ in order to obtain a mKdV equation at the minimum order of $\varepsilon$, see \cite{8}. 

By substituting the perturbed fields \eqref{3} into the coupled system \eqref{1} and \eqref{2} we obtain at the leading order of the perturbation, that is $O(\varepsilon^{0})$ the relations 
\begin{align}
  H_{0}^{x}=-M_{0}^{x}\, , \qquad M_{0}^{y}=-\lambda H_{0}^{y}\, , \qquad \mathrm{and} \qquad  M_{0}^{z}=-\lambda H_{0}^{z}\, . 
\end{align}
At the next order $O(\varepsilon)$, we obtain 
\begin{align}
H_{1}^{x}=-M_{1}^{x}\, , \qquad  M_{1}^{y}=-\lambda H_{1}^{y}\, ,\qquad  M_{1}^{z}=-\lambda H_{1}^{z}\, , 
\end{align}
and 
\begin{align}
-V\frac{\partial M_{0}^{x}}{\partial\zeta}&=\gamma(M_{0}^{y}H_{1}^{z}+M_{1}^{y}H_{0}^{z}-M_{0}^{z}H_{1}^{y}-M_{1}^{z}H_{0}^{y})\, ,\label{5a}\\
-V\frac{\partial M_{0}^{y}}{\partial\zeta}&=\gamma(M_{0}^{z}H_{1}^{x}+M_{1}^{z}H_{0}^{x}-M_{0}^{x}H_{1}^{z}-M_{1}^{x}H_{0}^{z})\, ,\label{5b}\\
-V\frac{\partial M_{0}^{z}}{\partial\zeta}&=\gamma(M_{0}^{x}H_{1}^{y}+M_{1}^{x}H_{0}^{y}-M_{0}^{y}H_{1}^{x}-M_{1}^{y}H_{0}^{x})\, .\label{5c}
\end{align}
However, it must be noted that from the linear relations in Eq. \eqref{5a} we get $M_{0}^{x}=0$ and then $H_{0}^{x}=0$.
By collecting terms at the next order $O(\varepsilon^{2})$ we obtain $H_{2}^{x}=-M_{2}^{x}$ and 
\begin{align}
\frac{\partial^{2}}{\partial\zeta^{2}}\left ( H_{2}^{y}-\frac{1}{\lambda}M_{2}^{y}\right) &=-\frac{\partial^{2}M_{0}^{y}}{\partial\zeta\partial\tau}\label{6a}\\
\frac{\partial^{2}}{\partial\zeta^{2}}\left ( H_{2}^{z}-\frac{1}{\lambda}M_{2}^{z}\right ) &=-\frac{\partial^{2}M_{0}^{z}}{\partial\zeta\partial\tau}\label{6b}\, ,
\end{align}
where $\lambda=(V^{2}-c^{2})/V^{2}$, as well as 
\begin{align}
  \begin{split}
-V\frac{\partial M_{1}^{x}}{\partial\zeta}&=Jf\left(M_{0}^{y}\frac{\partial^{2}M_{0}^{z}}{\partial\zeta^{2}}-M_{0}^{z}\frac{\partial^{2}M_{0}^{y}}{\partial\zeta^{2}}\right)+Jf_{\zeta}\left(M_{0}^{y}\frac{\partial M_{0}^{z}}{\partial\zeta}-M_{0}^{z}\frac{\partial M_{0}^{y}}{\partial\zeta}\right)\\
&+\gamma(M_{0}^{y}H_{2}^{z}+M_{1}^{y}H_{1}^{z}+M_{2}^{y}H_{0}^{z}-M_{0}^{z}H_{2}^{y}-M_{1}^{z}H_{1}^{y}-M_{2}^{z}H_{0}^{y})\, .
  \end{split}\label{7a}
\end{align}
At this order, the $y$ and $z$ components can be written by replacing the components of ${\bf{M}}$ and ${\bf{H}}$ cyclically. Then, after substituting the relations obtained at previous orders in Eq. \eqref{7a}, we end up with the reduced equation
\begin{align}
  \begin{split}
-V\frac{\partial M_{1}^{x}}{\partial\zeta} &=Jf\left(M_{0}^{y}\frac{\partial^{2}M_{0}^{z}}{\partial\zeta^{2}}-M_{0}^{z}\frac{\partial^{2}M_{0}^{y}}{\partial\zeta^{2}}\right)+Jf_{\zeta}\left(M_{0}^{y}\frac{\partial M_{0}^{z}}{\partial\zeta}-M_{0}^{z}\frac{\partial M_{0}^{y}}{\partial\zeta}\right) \\
&+\gamma\left[M_{0}^{y}\left(H_{2}^{z}-\frac{1}{\lambda}M_{2}^{z}\right)-M_{0}^{z}\left(H_{2}^{y}-\frac{1}{\lambda}M_{2}^{y}\right)\right]\, .
  \end{split}\label{8}
\end{align}
The corresponding equations for the $y$ and $z$ components are
\begin{align}
\frac{\partial M_{0}^{y}}{\partial\zeta} &=-\frac{(1-\lambda)\gamma}{V\lambda}M_{0}^{z}M_{1}^{x}\label{9a}\\
\frac{\partial M_{0}^{z}}{\partial\zeta} &=\frac{(1-\lambda)\gamma}{V\lambda}M_{0}^{y}M_{1}^{x}\label{9b}\\
-V\frac{\partial M_{1}^{y}}{\partial\zeta} &=\gamma\left[M_{0}^{z}\left(H_{2}^{x}-\lambda^{-1}M_{2}^{x}\right)+(\lambda^{-1}-1)M_{1}^{x}M_{1}^{z}\right]\label{9c}\\
-V\frac{\partial M_{1}^{z}}{\partial\zeta} &=\gamma\Big[(1-\lambda^{-1})M_{1}^{x}M_{1}^{y}-M_{0}^{y}\Big(H_{2}^{x}-\lambda^{-1}M_{2}^{x}\Big)\Big]\, .\label{9d}
\end{align}
Since $M_{0}^{x}=0$ from Eq. \eqref{5a}, we can also assume without loss of generality that the zeroth order field take the form \cite{8}
\begin{align}
{\bf{M}}_{0}=(0, \textmd{cos}(\theta), \textmd{sin}(\theta))\, ,\label{10}
\end{align}
where $\theta$ is a function of $(\zeta, \tau)$. 
Eq. \eqref{10} is therefore a spherical coordinate representation with $M_{0}^{x}=0$.
Using Eq. \eqref{10}, Eq. \eqref{9a} or \eqref{9b} can be written as
\begin{align}
M_{1}^{x}=\frac{V\lambda}{\gamma(1-\lambda)}\frac{\partial\theta}{\partial\zeta}\, , \label{11}
\end{align}
and substituting Eq. \eqref{10} and \eqref{11} in Eq. \eqref{8} yields the following perturbed modified KdV equation 
\begin{align}
  \begin{split}
\frac{\partial u}{\partial\tau}+\frac{3}{2}(A-\alpha f)u^{2}\frac{\partial u}{\partial\zeta}&+(A-\alpha f)\frac{\partial^{3} u}{\partial\zeta^{3}}=\alpha\Big\{f_{\zeta\zeta}\frac{\partial u}{\partial\zeta}\\
&+3f_{\zeta}\frac{\partial^{2} u}{\partial\zeta^{2}}-f_{\zeta\zeta\zeta}u+f_{\zeta}u^{3}+\frac{1}{2}\frac{\partial u}{\partial\zeta}\int_{-\infty}^{\zeta}f_{\zeta}u^{2}d\zeta'\Big\}\, ,
  \end{split}
\label{12}
\end{align}
where $u=\frac{\partial\theta}{\partial\zeta}$. For the homogeneous case, that is when $f=1$, Eq. \eqref{12} reduces to completely integrable modified KdV equation with $N$-soliton solutions \cite{43}
\begin{align}
\frac{\partial u}{\partial\tau}+\frac{3}{2}(A-\alpha)u^{2}\frac{\partial u}{\partial\zeta}+(A-\alpha)\frac{\partial^{3}u}{\partial\zeta^{3}}=0\, .\label{14}
\end{align}
For instance, the one soliton solution can be written as 
\begin{align}
u(\zeta,\tau)=2a\, \mathrm{sech}\left(a[\zeta-b\tau]\right )\, , \label{15}
\end{align}
for a constant $b$ and $a^{2}=b/(A-\alpha)$. Eq. \eqref{12} is therefore a perturbed integral modified KdV (PIMKdV) equation with inhomogeneities. 
The inhomogeneity will have the effect of modulating the shape of the soliton through which the magnetization is developed in the ferromagnet. 
This will be the subject of the rest of this paper.

\section{Effect of inhomogeneity on soliton propagation}\label{sec-4}

It has already been reported in \cite{44} that the solitons of the modified KdV obtained in the absence of damping and Heisenberg exchange correspond to the complete revolution of the magnetization vector about the propagation direction. 
In the integrable equation, the shape of the soliton is preserved throughout the propagation. 
However, the presence of damping and interactions such as Heisenberg and anisotropy modulates the shape and modifies the propagation of the solitons. 
In the present case, the inhomogeneity has a significant effect on the magnetization dynamics of the ferromagnet. 
The soliton responsible for the localization is deformed by the presence of inhomogeneities and in particular its structure and speed. 

A system with inhomogeneities is usually described by the variable coefficient nonlinear equations which still admit breathers, rogue waves in addition to that of solitons. 
Recently, in the context of the above systems, Yeping Sun \cite{45}, solved the Hirota equation with the spatially varying inhomogeneities, confirming the integrability of the equation. The author obtains the soliton, breather and rogue waves modes through the use of Darboux transformations. Some more interesting effects of inhomogeneities on the dynamical model may be found in \cite{46,47,48,49,50}. 
The PIMKdV Eq. \eqref{12} derived in the previous section is tedious to solve for the analytical solution and do not seem to be integrable, hence we rely on the numerical simulation for solving it, done using the python package Dedalus \cite{51}. 
This numerical scheme used in a spectral method based on $1024$ Fourier modes in space and a second order backward differentiation formula (BDF) time stepping scheme. 
Hence, all the simulations have periodic boundary conditions. 

\subsection{Periodic inhomogeneity}

\begin{figure}
\centering
\includegraphics[width=0.7\textwidth]{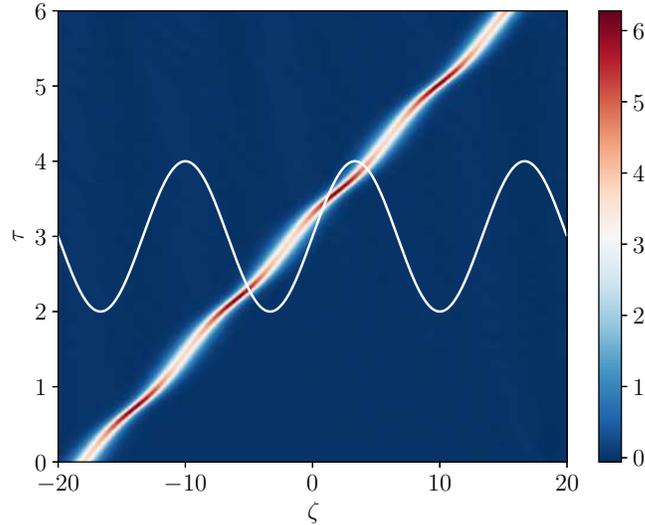}
\caption{In this figure, we illustrate the effect of a periodic inhomogeneity such as a sine function  (in white) on the propagation of a single soliton. 
The soliton oscillates as the sine function and constantly emits small radiative waves which will eventually destroy the original mKdV soliton. }
\label{fig:period}
\end{figure}

The first choice for the inhomogeneity is periodic and can arise from the periodic repetition of the lattice deformation. 
We use a model described by $f(\zeta)=0.2\, \sin(5\pi \zeta/L)$, for a domain $(-L, L)$ with domain size $L=20.0$ in Fig. \ref{fig:period}. 
The magnetization evolves in the form of solitons with periodic oscillations in its shape and speed as it traverses the inhomogeneity of the ferromagnet. 
The soliton constantly emits small radiative waves. 
This suggests that this system is not integrable and that the soliton will eventually evaporate. 
The amount of radiating waves is related to the steepness of the function $f$, as we will see in the next section with a steeper function. 

\subsection{Soliton propagation under a localised inhomogeneity}
\begin{figure}
\centering
\includegraphics[width=0.7\textwidth]{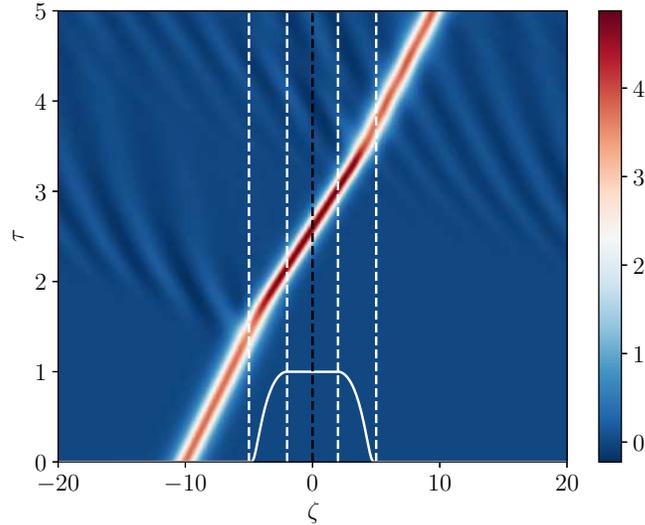}
\caption{This figure shows a one soliton colliding a bump function \eqref{f-bump}, represented in white at the bottom of the plot.
We observe a speedup of the soliton on the bump, and the creation of radiative waves, and the opposite effect on the other side of the bump.}
\label{fig:one}
\end{figure}

We now consider a localised inhomogeneity which forms a transition between two constant values of $f$ and is periodic. 
We use the bump function 
\begin{align}
    f(\zeta)=
    \begin{cases}
    0 & \text{if}~|\zeta|\geq l_1\\
    l_ae^{-1} & \text{if}~|\zeta|>l_2\\
    l_a e^{-\frac{1}{1- (\zeta-l_2)/\Delta l }^2} &\text{if}~l_2 <\zeta<l_1\\
    l_a e^{-\frac{1}{1- (\zeta+l_2)/\Delta l }^2} &\text{if}~-l_2 >\zeta>-l_1
    \end{cases}
    \label{f-bump}
\end{align}
where $l_1$ and $l_2$ are radii for the inner and outer regions and $l_a$ the amplitude of the bump. We also denote the size of the region where the function $f$ is non constant by $\Delta l=l_1-l_2$ for simplicity. We choose this function so that we can observe the effect of a non-constant function, a non-vanishing constant function and its value going back to $0$. This plateau allows us to demonstrate that for a small perturbation, the original magnetic soliton sees its speed and shape affected. We can directly compute the speed of the soliton on the plateau, as the inhomogeneity is constant, thus the equation of motion corresponds to a mKdV equation with different coefficient. 
From the value of $f_\mathrm{max}= l_a e^{-1}$ as 
\begin{align}
    V = \frac{V_{0}}{1-\alpha f_\mathrm{max}}\, ,
\end{align}
for an initial speed $V_{0}$ (for vanishing $f$), and for the choice $A=1$. This can be seen as the right hand side only depends on the derivatives of $f$ and the left hand side will have constant coefficients. After passing the bump, the soliton returns to its original speed $V_{0}$. We illustrates this in Figure~\ref{fig:one} with parameters $A=1, \alpha=0.2$ and initial soliton with amplitude of $3.5$. 
The soliton can see its speed increased or decreased if $\alpha f>0$ or $\alpha f<0$ in the region of non-constant $f$. 
From this numerical simulations, we can also observe the production of some dispersive radiations when $f$ is non-constant, or larger amplitude than for the periodic inhomogeneity. 

\subsection{Soliton-breather transition}

We now study an intriguing feature of this equation, when two mKdV solitons with opposite polarities (one positive and one negative) and with different speeds collide on a localised inhomogeneity $f$, such as a Gaussian, or sech bump, which we will call a wall. 
We chose it as 
\begin{align}
    f(\zeta) = 2 \beta \, \mathrm{sech}(2\zeta)\, , 
\label{f_sech}
\end{align}
where $\beta$ parametrises the height of the wall. 
The two solitons have the standard form of $\pm \sqrt{B}\, \mathrm{sech}(\sqrt{B}\zeta)$, where $B= 3.5$ and $6$ for the positive and negative solitons respectively.
We used periodic boundary conditions and the initial condition has solitons separated enough from the wall and between them such that the $2$-soliton solution of the perturbed mKdV is well-approximated by the sum of these two $1$-soliton solution of the un-perturbed mKdV, given in equation \eqref{f_sech}. 

\begin{figure}
\centering
\subfigure[No wall, $\beta=0.0$ in \eqref{f_sech}.]{\includegraphics[width=0.48\textwidth]{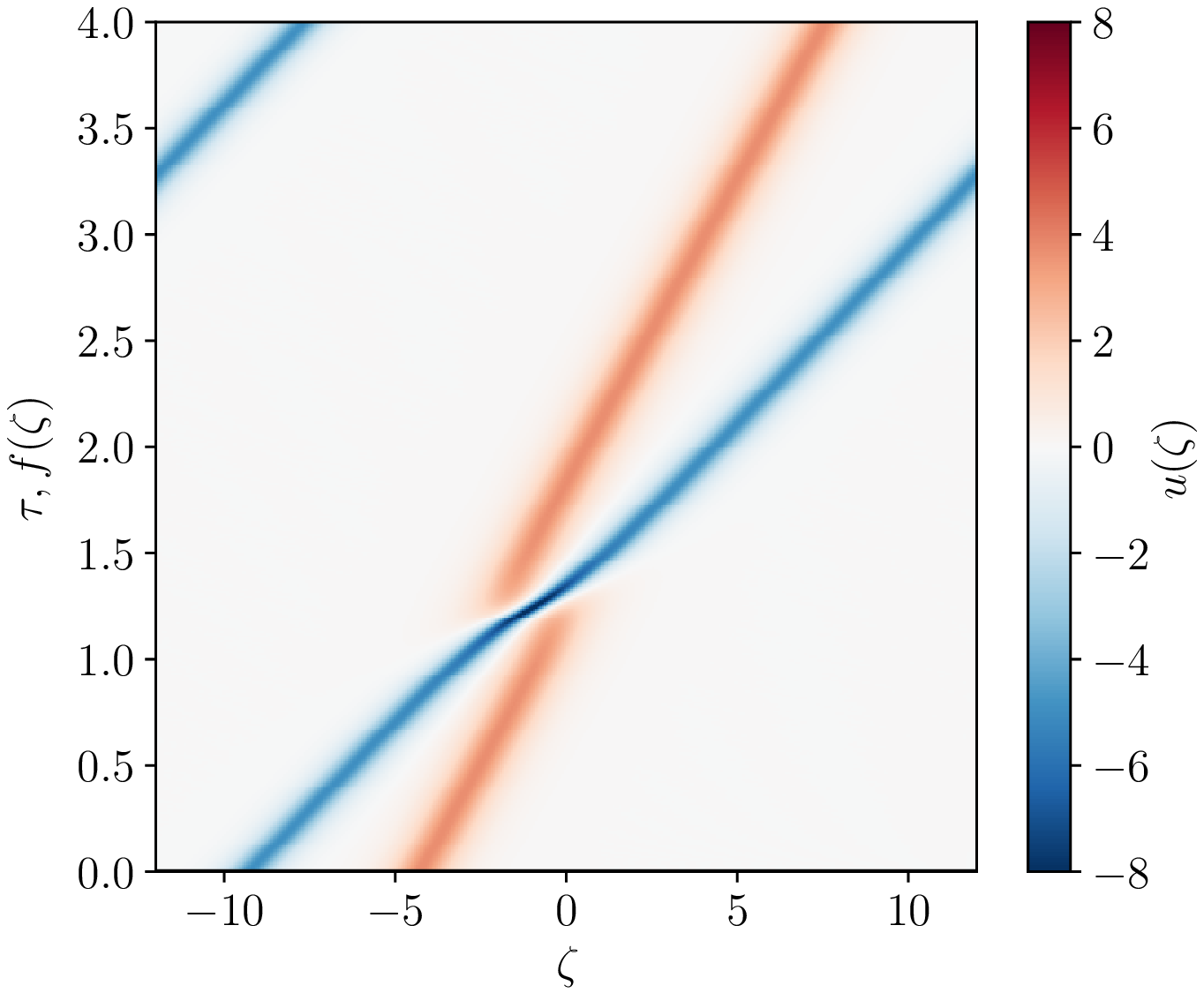}\label{fig:breather_0}}
\subfigure[Small wall, $\beta=0.3$ in \eqref{f_sech}.]{\includegraphics[width=0.48\textwidth]{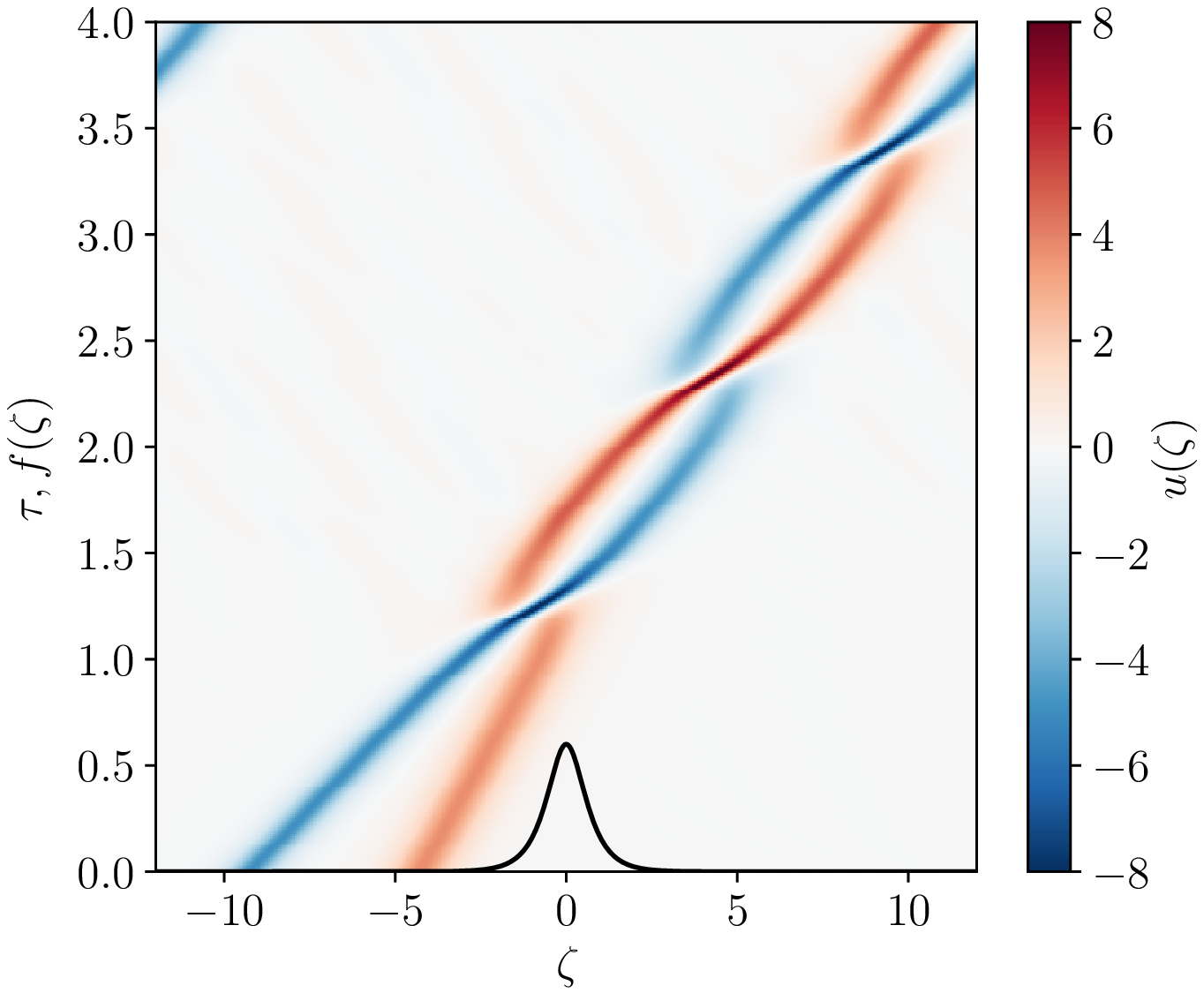}\label{fig:breather_1}}
\subfigure[Medium wall, $\beta=0.95$ in \eqref{f_sech}.]{\includegraphics[width=0.48\textwidth]{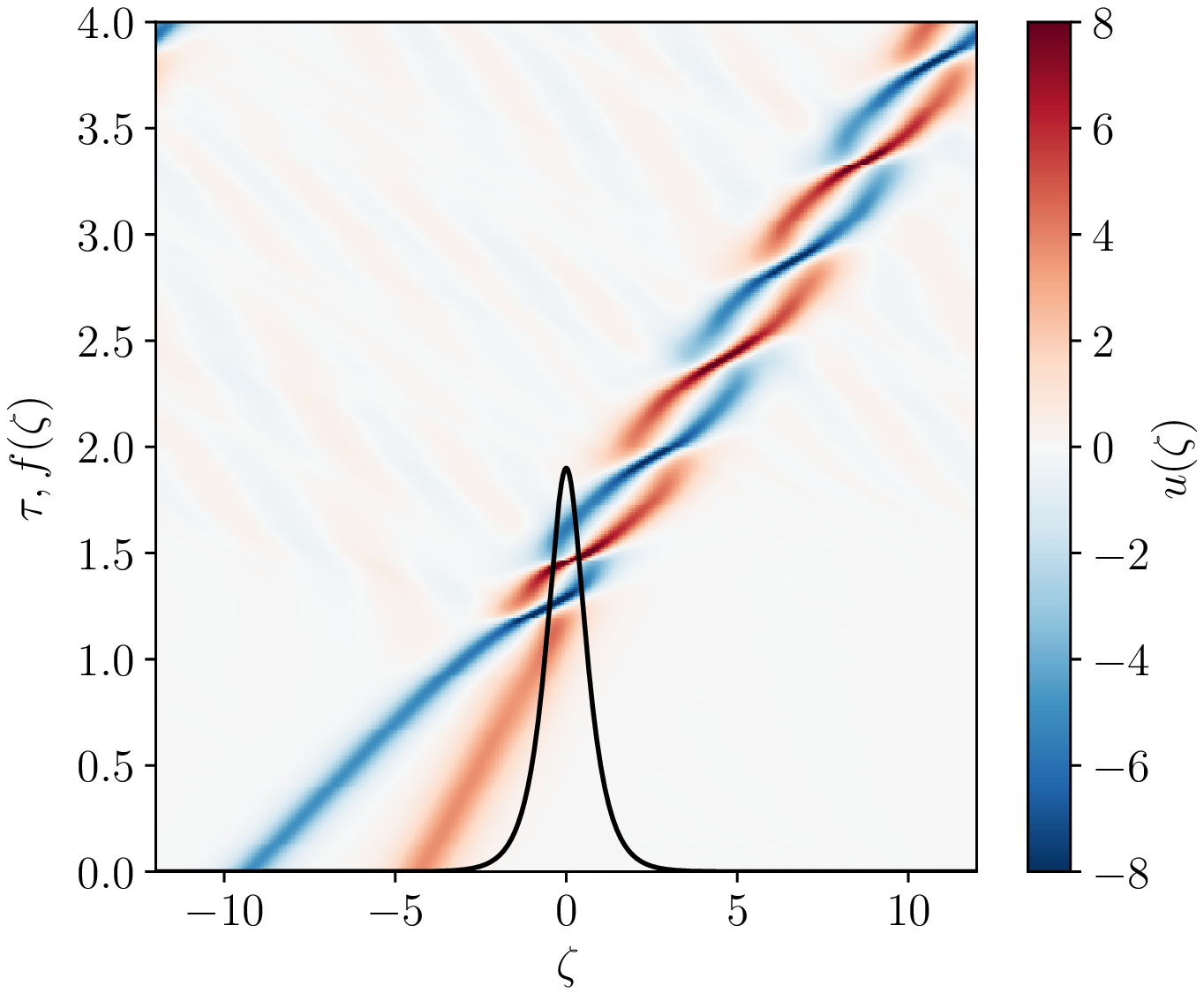}\label{fig:breather_2}}
\subfigure[Large wall, $\beta=1.28$ in \eqref{f_sech}.]{\includegraphics[width=0.48\textwidth]{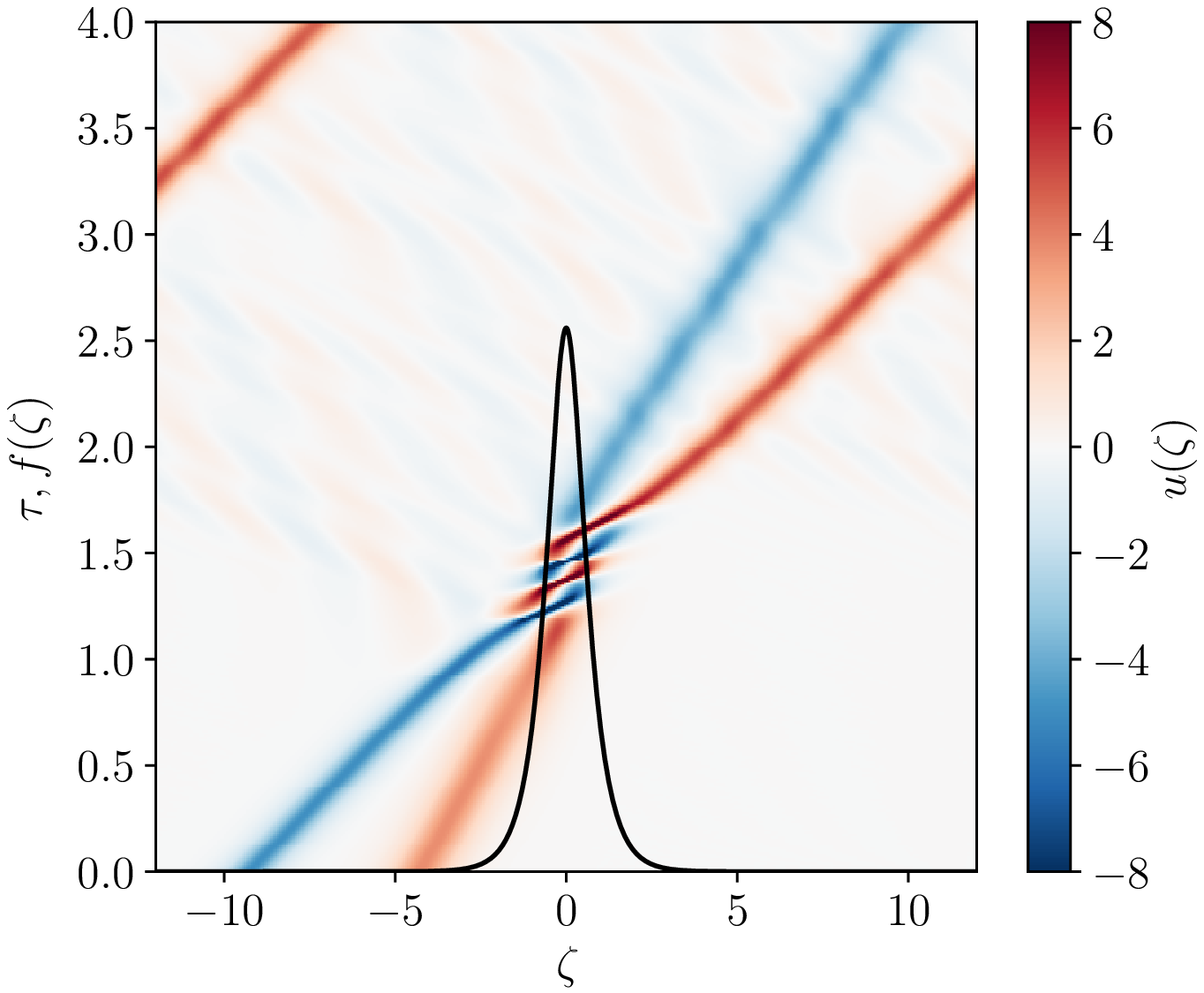}\label{fig:breather_3}}
\caption{
  In this figure, we illustrate the three types of collision of solitons on a wall function, shown in black at the bottom of the plots, and a normal collision in panel \ref{fig:breather_0} for comparison.  
  On panel \ref{fig:breather_1}, the presence of a small wall creates a breather with a small frequency, and numerically stable, see \cite{52} for an analytical expression of the breather. 
  For a different wall amplitude, the simulation in panel \ref{fig:breather_2} shows that the breather created on the wall may not persist and separate in two solitons, with different amplitudes than the original ones. }
\label{fig:breather}
\end{figure}

\begin{figure}[htpb]
  \centering
\includegraphics[width=0.6\textwidth]{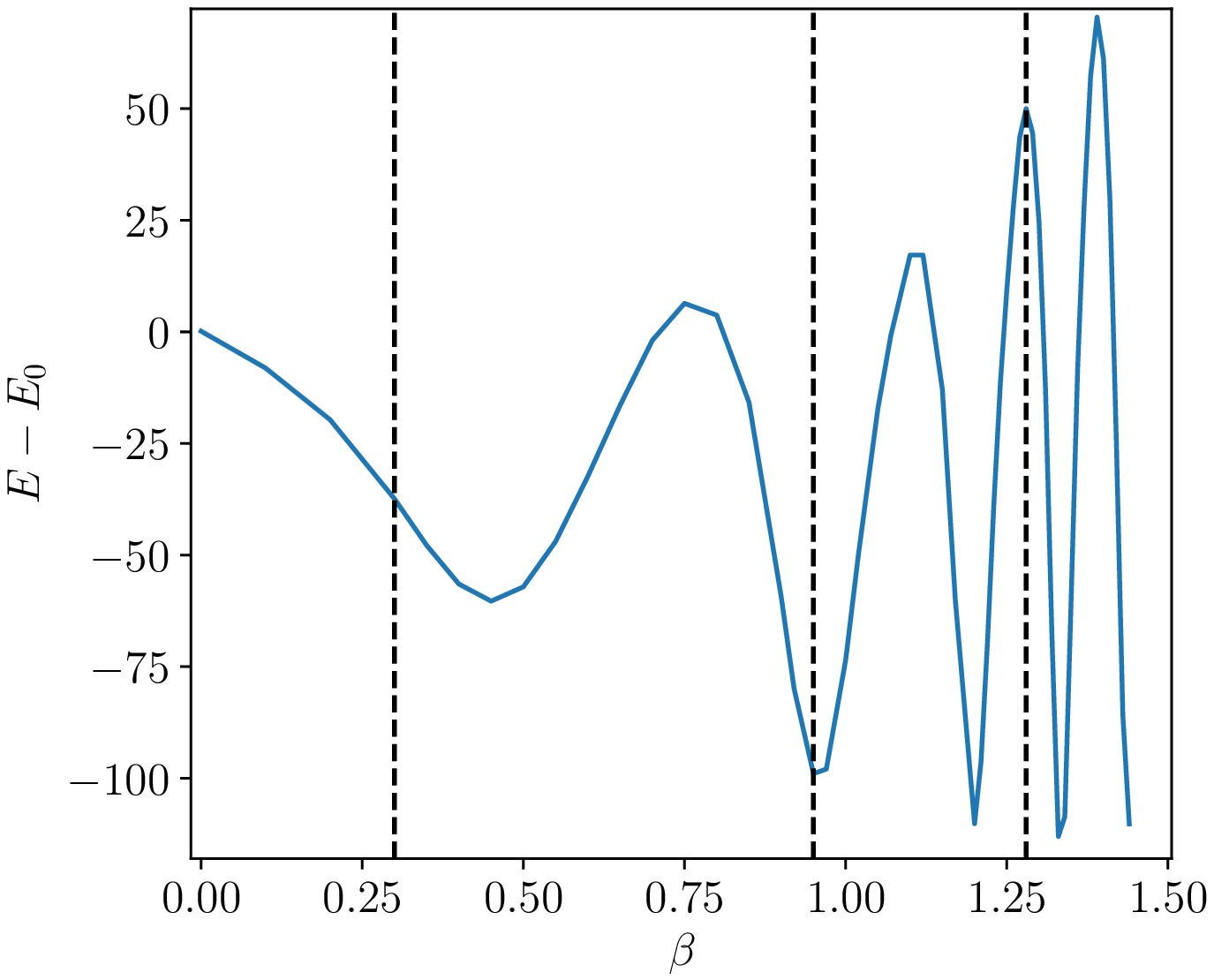}
  \caption{
    In the figure, we plot the difference of the energy of the outgoing solitons or breather with the initial energy of many simulations with various values of $\beta$.  This shows a periodic behaviour of the two solitons or breather outgoing solutions of figure \ref{fig:breather}, the vertical lines corresponds to the three solutions of figure \ref{fig:breather}.  }
\label{fig:E-E0}
\end{figure}

Depending on several parameters, such as the amplitude and profile of the wall, the speed of the solitons and the exact position of the collision with respect to the wall, the collision undergoes different behaviours. 
To analyse these different behaviours, we run numerical simulations with an increasing amplitude of the wall via the parameter $\beta$ and we display the results in figure \ref{fig:breather}.
The effect of the wall on the soliton collision is to create a bound state of the two solitons and then release it as two single solitons, or as a single breather, see \cite{52} for their analytical form. 
The breathers and solitons have properties which depend continuously on $\beta$ with a certain periodicity. 
This is shown in figure \ref{fig:E-E0} where we compare the energy of the outgoing solitons or breather, with the energy of the two incoming solitons, where the energy is defined as 
\begin{align}
  E = \int \left (\frac18 u(\zeta)^4 - \frac12 u_\zeta(\zeta)^2 \right )d\zeta \, . 
\end{align}
We observe a clear periodicity of the type of outgoing waves out of the collision on the wall, which is accelerating for larger $\beta$. 
We could not go beyond these values, as the numerical solution was becoming less reliable. 
Indeed, for a large and steep wall, the integro-differential equation requires a precise numerical scheme, which is out of the scope of this first investigation. 

Nevertheless, we can give a tentative explanation for the mechanism of the creation of these structures. 
Recall that the effect of an increasing inhomogeneity function $f(\zeta)$ is to increase the speed of the soliton if it has the same sign as the soliton. 
In the collision of two solitons on a wall, the negative slow soliton is accelerated whereas the fast positive soliton is decelerated. 
This deceleration occurs only after acceleration of the slow soliton, as it attains the wall later. 
Hence, depending on how much and when the change of speed happens, the two solitons collide with a smaller speed difference than without the wall. 
The interaction of solitons contributes to exchanging energy between them, supplemented by this continuous effect of an increasing function $f$. 
If the parameters of the wall and initial condition are adjusted appropriately, the two solitons may merge to form a breather or separate with different speeds, but not necessarily the original ones. 

\section{Conclusion}\label{sec-5}

To conclude, we performed a multiscale analysis on the Landau-Lifshitz equation coupled to Maxwell's equations for the propagation of electromagnetic waves in the deformed ferromagnetic medium. 
The LLM model is reduced to the new perturbed integral modified KdV equation at the second order approximation of the perturbation for the particular choice $m=1$ and $n=1$. 

In order to study the localised evolution of the magnetization component driven by the inhomogeneities, we solved the perturbed integral modified KdV equation with the help of numerical simulations. 
In particular, we simulated two solitons of opposite polarities and made them interact with a localised inhomogeneity. 
As expected, the presence of inhomogeneity delivers dissipation into the system which supports the soliton excitation, but can also inject energy into it. 
They, in turn, deform locally and regenerate into a breather after the encounter with the localised inhomogeneity. 
The soliton deformation for different amplitudes of the inhomogeneity is reported and shown to create different breathers, with a positive or a negative direction of propagation with respect to the direction of propagation of the two solitons.  
We thus demonstrated here that such breathers can emerge spontaneously from the collision of solitons in an inhomogeneous medium. 
Further detailed numerical investigation of this phenomenon should be undertaken as this work is only a first exploration of this phenomenon. 

We finally want to discuss the origin of the inhomogeneities described in this work. 
They can represent the presence of foreign atoms such as drugs, mutants and dyes in a particular site of the DNA sequence \cite{53}. 
In this context, the localised inhomogeneities modify the shape of the soliton which usually is a kink in DNA dynamics. 
It must be noted that localised inhomogeneities favour the denaturation process, the local opening modes in the DNA molecule as well as the propagation of phonon modes. 
In another context, the presence of magnetic inhomogeneities significantly affects the damping time and ferromagnetic resonance (FMR) linewidth broadening. The variation of the FMR linewidth broadening depends on the strength of the inhomogeneities \cite{54}. These FMR linewidth experiments exploit the magnetization precession from one stable configuration to other within a minimum time duration at the order of few nanoseconds. This is an important issue for the data storage and other magnetic device applications.  
Thus, from the reports published earlier on the effect of localised inhomogeneities, we believe that our present study on the inhomogeneities in the magnetic system may help to enhance the technological aspects of constructing new devices based on magnetic materials. 

\section*{Acknowledgement}
The authors are grateful to the referees whose reports helped improve the exposition of these results. 
AA acknowledges the EPSRC through award EP/N014529/1 funding the EPSRC Centre for Mathematics of Precision Healthcare.


\section*{References}

\begin{thebibliography}{99}
\bibitem{1}Th Gerrits, H.A.M. van den Berg, J. Hohlfeld, L. B${\ddot{a}}$r, Th. Rasing, Nature 418 (2002) 509.
\bibitem{2}M.G. Cottam, Linear and Nonlinear  Waves in Magnetic Films and Superlattices, Singapore, World Scientific, 1994.
\bibitem{3}W.J. Caspers, Spin Systems, Singapore, World Scientific, 1993.
\bibitem{4}A.I. Akhiezer, V.G. Baryakhtar, S.V. Peletminskii,  Waves, Amsterdam, North-Holland, 1968.
\bibitem{5}B.A. Malomed, Soliton Management in Periodic Systems, New York, Springer, 2006.
\bibitem{6}J.L. Simonds, Phys. Today 48(4) (1995) 26.
\bibitem{7}R.F. Soohoo, Theory and Applications of Ferrites, London, Prentice Hall, 1960.
\bibitem{8}I. Nakata, J. Phys. Soc. Japan 60 (1991) 77.
\bibitem{9}I. Nakata, J. Phys. Soc. Japan 60 (1991) 3976.
\bibitem{10}H. Leblond, J. Phys. A: Math. Gen. 36 (2003) 1855.
\bibitem{11}H. Leblond, J. Phys. A: Math. Gen. 28 (1995) 3763.
\bibitem{12}H. Leblond, M. Manna, J. Phys. A: Math. Gen. 26 (1993) 6451.
\bibitem{13}H. Leblond, M. Manna, J. Phys. A: Math. Theor. 41 (2008) 185201.
\bibitem{14}H. Leblond, J. Phys. A: Math. Gen. 29 (1996) 4623.
\bibitem{15}H. Leblond, J. Phys. A: Math. Gen. 32 (1999) 7907.
\bibitem{16}M. Daniel, V. Veerakumar, R. Amuda, Phys. Rev. E 55 (1997) 3619.
\bibitem{17}V. Veerakumar, M. Daniel, Phys. Rev. E 57 (1998) 1197.
\bibitem{18}M. Daniel, V. Veerakumar, Phys. Lett. A 302 (2002) 77.
\bibitem{19}K. Nakamura, T. Sasada, Phys. Lett. A 48 (1974) 321.
\bibitem{20}M. Lakshmanan, Phys. Lett. A 61 (1977) 53.
\bibitem{21}L.A. Takhtajan, Phys. Lett. A 64 (1977) 235.
\bibitem{22}M. Daniel, M.D. Kruskal, M. Laksmanan, K. Nakamura, J. Math. Phys. 33 (1992) 771.
\bibitem{23}T. Masuda, A. Zheludev, B. Roessli, A. Bush, M. Markina, A. Vasilev, Phys Rev B 72 (2005) 014405.
\bibitem{24}W.Z. Zhao, Y.Q. Bai, K. Wu, Phys Lett A 352 (2006) 64.
\bibitem{25}D. Giridharan, P. Sabareesan and M. Daniel, Phys. Rev. E 94 (2016) 032222.
\bibitem{26}Francis T. Nguepjouo, Victor K. Kuetche, Timoleon C. Kofane, Phys. Rev. E 89 (2014) 063201.
\bibitem{27}Victor K. Kuetche, Francis T. Nguepjouo, Timoleon C. Kofane, Chaos, Soliton and Fractals 66 (2014) 17.
\bibitem{28}Hermann T. Tchokouansi, Victor K. Kuetche, Timoleon C. Kofane, Chaos, Soliton and Fractals 86 (2016) 64.
\bibitem{29}Victor K. Kuetche, Hermann T. Tchokouansi, Timoleon C. Kofane, Jour. Magn. Magn. Mater. 374 (2015) 1.
\bibitem{30}Victor K. Kuetche, Jour. Magn. Magn. Mater. 398 (2016) 70.
\bibitem{31}M. Saravanan, Phys. Lett. A 378 (2014) 3021.
\bibitem{32}J.D. Jackson, Classical Electrodynamics, New York, Wiley Eastern, 1993.
\bibitem{33}Xiaotong Liu, Xuelin yong, Yehui Huang, Rui Yu, Jianwei Gao, Commun. Nonlinear Sci. Numer. Simulat. 29 (2015) 257.
\bibitem{34}G. Duerr, R. Huber, D. Grundler, J. Phys.: Condens. Matter 24 (2012) 024218.
\bibitem{35}R. Balakrishnan, J. Phys. C: Solid State 15 (1982) L1305.
\bibitem{36}L.N. Bulaevskii, A.V. Zvarykina, YuS. Karimov, R.B. Lyobovskii, I.F. Shchegolev, Sov. Phys.-JETP 35 (1972) 384.
\bibitem{37}J.C. Bonner, H.W.J. Blote, H. Beck, G. Muller, Physics in One Dimension ed., J. Bernasconi and T. Schneider, New York, Springer-Verlag, 1981.
\bibitem{38}H. Leblond, J. Phys. A: Math. Gen. 34 (2001) 9687.
\bibitem{39}A. Jeffrey, T. Kawahara, Asymptotic Methods in Nonlinear Wave Theory, Boston, Pitman Advanced Publishing Program, 1982.
\bibitem{40}Saravanan M, Emmanuel Yomba, Chaos Solitons and Fractals 103 (2017) 139.
\bibitem{41}M. Saravanan, Phys. Rev. E 92 (2015) 012923.
\bibitem{42}Ch. Kittel, Phys. Rev. 73 (1948) 155.
\bibitem{43}M. Wadati, J. Phys. Soc. Jpn. 32 (1972) 1681.
\bibitem{44}H. Leblond, J. Phys. A: Math. Gen. 33 (2000) 8105.
\bibitem{45}Yeping Sun, Xuelin Yong and Jianwei Gao, Chaos, Soliton and Fractals, 77 (2015) 101.
\bibitem{46}J.L. Helm, T.P. Billiam and S.A. Gardiner, Phys. Rev. A (2012) 053621. 
\bibitem{47}J. Cuevas, P.G. Kevrekidis, B.A. Malomed, P. Dyke and R.G. Hulet, New J. Phys. 15 (2013) 063006.
\bibitem{48}Z.-Y. Sun, P.G. Kevrekidis and P. Kruger, Phys. Rev. A 90 (2014) 063612. 
\bibitem{49}Z.-Y. Sun, P.G. Kevrekidis and P. Kruger, Phys. Rev. A 94 (2016) 063645. 
\bibitem{50}Z.-Y. Sun, Y.-T. Gao, Y. Liu and X. Yu, Phys. Rev. E 84 (2011) 026606.
\bibitem{51}K. J. Burns, G. M. Vasil, J. S. Oishi, D. Lecoanet, B. P. Brown, and E. Quataert, Astrophysics Source Code Library (2016).
\bibitem{52}A. V. Slunyaev and E. N. Pelinovsky, Phys. Rev. Lett. 117 (2016) 214501.
\bibitem{53}M. Daniel and M. Vanitha, Phys. Rev. E 84 (2011) 031928.
\bibitem{54}R.D. McMichael, D.J. Twisselmann and Andrew Kunz, Phys. Rev. Lett. 90 (2003) 227601-1. 
\end{thebibliography}
\end{document}